\documentclass[print]{aa}
\usepackage[colorlinks,linkcolor=blue,anchorcolor=green,citecolor=blue]{hyperref}

\usepackage{graphicx}
\usepackage{txfonts}
\usepackage{units}

\begin{document}

    \title{The near-infrared radiation background, gravitational wave background and star
formation rate of Pop III and Pop II during cosmic reionization}
    \author{Y. P. Yang\inst{1,2}, F. Y. Wang\inst{1,2} \and Z. G. Dai\inst{1,2}}

    \institute{School of Astronomy and Space Science, Nanjing University, Nanjing
210093, China\\
        \and
        Key Laboratory of Modern Astronomy and Astrophysics (Nanjing University), Ministry of Education, Nanjing 210093, China\\
        \email{fayinwang@nju.edu.cn}}


\titlerunning{NIRB, SBGWs and SFR of Pop III and Pop II during cosmic reionization}
\authorrunning{Yang, Wang \& Dai}

\abstract
  {The transition from Population III
(Pop III) to Population II (Pop II) stars plays an important role in the universe history. Due to a huge
amount of ionizing photons generated by Pop III stars, they begin to ionize the intergalactic medium (IGM) at
the early stage of reionization. Meanwhile, the feedback
from reionization and metal enrichment changes the evolution of different populations. The near-infrared radiation background (NIRB)
and the background of gravitational waves (SBGWs) from these early stars will provide important information about the transition form Pop III to Pop II stars.}
{In this paper, we obtain the NIRB
and SBGWs from the early stars, which are
constrained by the observation of reionization and star formation rate.}
{We study the transition from Pop III to Pop II
stars via the star formation model of different population, which
takes into account the reionization and the metal enrichment evolution. We calculate the
two main metal pollution channels arising from the supernova-driven protogalactic outflows and ``genetic channel''. We obtain
the SFRs of Pop III and Pop II and their NIRB and SBGWs radiation.}
{We predict that the upper limit of metallicity in metal-enriched
IGM (the galaxies whose polluted via ``genetic channel'') reaches
$Z_{\rm crit}=10^{-3.5}Z_{\odot}$ at $z\sim13$ ($z\sim11$), which is
consistent with our star formation model. We constrain on the SFR of
Pop III stars from the observation of reionization. The peak
intensity of NIRB is about $0.03-0.2~
\unit{nW}\unit{m}^{-2}\unit{sr}^{-1}$ at $\sim 1 \unit{\mu m}$ for
$z>6$. The prediction of NIRB signal is consistent with the
metallicity evolution. We also obtain the gravitational wave
background from the black holes formed by these early stars. The
predicted gravitational wave background has a peak amplitude of
$\Omega_{GW}\simeq8\times10^{-9}$ at $\nu=158~\unit{Hz}$ for Pop II
star remnants. However, the
background generated by Pop III.2 stars is much less than Pop II stars, with a peak amplitude of $\Omega_{GW}\simeq1.2%
\times10^{-11}$ at $\nu=28~\unit{Hz}$. The background of Pop III.1
shifts to lower frequencies, and the amplitude of $\Omega _{GW}$ for
Pop III.1 stars shows a minimum value at $\nu\simeq 10~\unit{Hz}$,
due to the lack of gravitational wave signals from the stars with $%
140~M_{\odot}<M_\ast<260~M_{\odot}$.}
{}
\keywords
{cosmology: theory -- diffuse background -- intergalactic medium -- gravitational waves -- stars: early type}
\maketitle

\section{INTRODUCTION}
The first stars, known as Population III (Pop III), form at the end
of the cosmic dark age and open the reionization era. Pop
III stars can have different characteristics compared to Pop I/II
stars. Some authors have stated that Pop III can be very
massive, on the order of $500-600M_{\odot }$
\citep{abe02,omu01,omu03,bro04a}, while others have claimed that the
masses are lower, on the order of $30-60M_{\odot }$ or $10M_{\odot}$
\citep{hos11,hos12,sta12}. The initial conditions for Pop III star
formation are entirely determined by the basic parameters of the
cosmology \citep{teg97,yos03,bro09}. According to the popular
\textquotedblleft bottom-up\textquotedblright\ hierarchical
structure formation in the standard cold dark-matter (CDM) model,
the first stars form in a sufficient amount of cold dense gas in a
dark matter halo with mass $\gtrsim 10^{6}\sim 10^{8}M_{\odot }$,
which collapses at redshift $z\sim 20-30$ \citep{bro99,bro02,abe02}.
At first, these stars contain no elements heavier than helium, viz.,
metal free, so-called Pop III.1. The second generations
of stars originate from the environment influenced by
earlier star formation, defined as Pop III.2, which exist in a
pre-ionized region and are affected by previous generations of
stellar radiation due to photo-ionization of ambient neutral
hydrogen of Pop III.1 stars. Because of supernova (SN) explosions
and stellar winds, heavier elements are introduced into the
intergalactic medium (IGM), gradually increasing its metallicity. On the other hand, some new galaxies inherit the metals from their low-mass progenitors.
After the metal elements enriching to a
certain threshold $Z_{\mathrm{crit}}$, the Pop I/II stars form %
\citep{bro01,bro03,sch02a,sch06,mac03}.

There has not been any direct observation of Pop III stars.
Fortunately, there are some observations to constrain the early
stellar population and reionization, e.g. the near-infrared
radiation background (NIRB), Gunn-Peterson trough
\citep{fan06} and the background of gravitational waves
\citep{mar09,per10}. NIRB contains some important information of early
universe, which arises from accumulated emission from the early
galaxy populations with a large range of redshifts. Because photons
lose energy in cosmic expansion, the optical and UV
radiation from the early stars during reionization should leave a
signature in the extragalactic background light at NIR bands with
wavelength less than a few micrometers. Although most high-redshift
galaxies are below the limiting magnitude of current detectors, we
could observe their redshifted cumulative spectrum,
which would be present in any background emission in the NIR band %
\citep{bar00,sal06,wyi06,kis09,bou10,rob10,fer11,mun11}. On the
other hand, since Pop II galaxies at $z\sim6$ are already observed
with Hubble Space Telescope (HST) and will be in a short time
studied in great details with James Webb Space Telescope (JWST), by
studying the NIRB intensity, we can find many properties of the
early stars, such as the star formation rate of high-redshift stars
whose photons are redshifted, the clustering properties (by studying
the anisotropy power spectrum of the NIRB), the classification of
Pop III stars (by studying the number of the Lyman alpha bumps), the
transition from Pop III to Pop II stars (by studying the evolution
of Lyman bumps), the reionization and metal enrichment (as we
discussed in this paper). However, determining the contribution of
these early stars at high redshifts from observations is a hard
task. There are two reasons: first, low redshift and very faint
galaxies make a significant contribution to NIRB, which is not
measured accurately. Second, the extragalactic NIRB is hard to
distinguish from the brighter zodiacal foregrounds in the local
matter, (e.g. interplanetary dust within the solar system), and the
stars and the interstellar medium (ISM) of our Galaxy
\citep{hau01,kas05a}, which is two to three orders of magnitude
larger than $10~\unit{nW}\unit{m}^{-2}\unit{sr}^{-1}$. Recent works
suggest that the intensity of NIRB cannot be larger than a few
tenths $\unit{nW}\unit{m}^{-2}\unit{sr}^{-1}$ %
\citep{kne04,ste06,fra08,gil09,gil12,fin10,kne10,ino13}. Except for mean
intensity, the anisotropy power spectrum of NIRB could provide direct
information of these early stars \citep{coo04b,kas04,fer10,coo12}.
The Pop III epoch contains less projected volume than the ordinary
galaxy populations (e.g. Pop II stars), leading to larger relative fluctuations, and their
anisotropy power spectrum represents the clustering behavior of sources in the IR regime.
However, for the ordinary galaxies (containing Pop II stars), their angular power spectrum is nearly a power law over a wide range of angular scales \citep{coo04b,coo12}.

The UV photons ($\lambda \sim 1000\mathrm{{\mathring{A}}}$)
produced at $z=9$ during reionization will be redshifted to the NIR
band with wavelength $\lambda \sim 1\unit{\mu m}$. Thus the NIRB
would depend on the evolution of the early stars during
reionization.  So far, many works have obtained some information about reionization, e.g.
the optical depth for electron scattering, the end of reionization redshift, and the escape fraction of ionizing photons, which can be used to constrain the SFRs of the early stars\citep{yu12,wan13}, and further constrain the
intensity of NIRB and SBGWs. Recently, there are many studies to calculate the
contribution of stars and galaxies responsible for reionization and
NIRB
\citep{kas02,san02,sal03,coo04a,coo04b,kas04,kas05a,kas05b,mad05,
mag03,fer06,kas07,tho07a,tho07b,fer10,coo12,fer12,
kas12,fer13a,fer13b,yue13}.
As a crucial part of the entire story of the cosmic history,
reionization reflects the transformation of neutral hydrogen in the
IGM into an ionized state, which is due to ionizing photons
generated by the early stars and galaxies. In the process of the
early star formation, the IGM is enriched with metals that
are dispersed by the first SNe and stellar winds. This process might
reflect the evolution of the stellar populations, although
it is not understood when the transition from Pop III to Pop II
takes place, due to an uncertainty in the \textquotedblleft critical
metallicity\textquotedblright\ $Z_{crit}$
\citep{bro01,jap09a,jap09b}, and the redshift evolution of the IGM
metallicity \citep{bar98,ferr00b,gre06} and some other metal pollution channels, e.g. ``genetic channel'' \citep{sch06b,tre09}.
With more Pop II star forming (although fewer ionizing photons per stellar), there are more
ionizing photons to ionize the IGM, and the ionized bubbles
gradually overlap, allowing the mean free path of ionizing
photons to increase rapidly. The average volume fraction of ionized
hydrogen in IGM increase rapidly. At last the entire universe is
almost completely ionized \citep{bar01,rob10,bro13}.

In this paper, We consider two main channels of metals pollution:
the supernova-driven protogalactic outflows \citep{ferr00b,fur05}
and the ``genetic channel'' \citep{sch06b,tre09}. The former assumed
that metals are enriched via supperbubbles resulting from supernova
explosion in protogalaxies, and the latter suggested that the new
galaxies inherit metals from the lower mass progenitor galaxies.
Moreover, we constrain the SFR through the observations and
reionization, e.g. optical depth for electron scattering as measured
by WMAP and Planck, and the redshift of the end of reionization with
a possible range from 5 to 10. During reionization, the metal
elements enrich in IGM. Pop III.1 stars originate from a freshly
collapsed halo, and Pop III.2 stars exist in a pre-ionized
metal-free region and they form via hydrogen deuteride cooling. Pop
II stars originate from the dark matter halos that have been
polluted by metal enrichment. Thus the SFRs of these early stars
would depend on the the hydrogen reionization fraction and pristine
fraction \citep{gre06,wan09b,sou11}. Following \citet{fer06}, this
NIRB consists of several contributions: the continuum emission from
stars themselves, the series of recombination lines, the free-free
and free-bound continuum emission from ionized gas or nebula, and
the two-photon emission. Based on the constraint on the SFRs of the
early stars, we can obtain the contribution of the first stars to
the high-redshift NIRB during reionization, which is lower than the
current observation  (the total luminosity $1\sim
10~\unit{nW}\unit{m}^{-2}\unit{sr}^{-1}$) %
\citep{san02,sal03,coo04a,kne04,kas05b,ste06,kas07,fra08,gil09,fin10,kne10,gil12,kas12,ino13}, due to the foreground pollution.
On the other hand, as pointed out by \citet{fer13b}, the Lyman $%
\alpha $ emission from Pop III stars at high redshift could result in \textquotedblleft bump\textquotedblright\ in spectrum of NIRB. This shape of Lyman
$\alpha$ \textquotedblleft bump\textquotedblright\ is determined by the transition from Pop III to Pop II stars.

In addition, these early stars are predicted to collapse
into black holes (except in the mass range $140-260M_\odot$, which
die as pair-instability supernova (PISN) \citep{heg02}), and
expected to be the sources of the stochastic background of
gravitational waves (SBGWs)
\citep{sch00,buo05,san06,suw07,mar09,per10}. Based on the SFRs, we
calculate the SBGWs that were produced by these early stars,
including Pop II, Pop III.2 and Pop III.1 stars.  At present, some gravitational wave interferometers
are operating in the frequency of $10-3000~\unit{Hz}$, e.g. VIRGO,
Laser Interferometer Gravitational-Wave Observatory (LIGO). In the
future, next-generation gravitational wave detection will open a
lower frequency window, e.g. Laser Interferometer Space Antenna
(LISA) covering the frequency range $10^{-4}-0.1~\unit{Hz}$, and Big
Bang Observer (BBO) operating in the range $0.01-10~\unit{Hz}$.  The
signals of gravitational waves will open a new window for the study
of the cosmic transition from Pop III to Pop II stars.

The structure of this paper is organized as follows. In Section 2,
we outline the stellar models, including the properties of distinct
populations and their SFRs during the era of reionization. In
Section 3. we give our reionization model by considering the
transition from Pop III to Pop II. In Section 4, we calculate the
spectrum of NIRB, which is contributed by the early stars before the
end of reionization. In Section 5, the gravitational wave spectra
produced by Pop III and Pop II stars are presented. In Section 6,
discussions and conclusions are given. In the whole paper, we assume
a flat $\Lambda $CDM model with $\Omega _{m}=0.27$, $\Omega
_{\Lambda }=0.73$, and $H_{0}=71$ km s$^{-1}$ Mpc$^{-1}$.

\section{THE STELLAR MODELS}

In this section, we consider the emissions from early stars during
reionization. Following \citet{fer06}, we calculate the emission
from two stellar populations. First, Pop III stars can be divided
into the first generation stars (Pop III.1), which their formation
only depends on initial conditions of the early universe, and the
second generation stars (Pop III.2), which exist in a pre-ionized
region and are affected by previous generations of stellar radiation
due to photo-ionization of ambient neutral hydrogen of Pop III.1
stars. They form via hydrogen deuteride
cooling \citep{joh06} with less masses and extremely poor metallicity $%
Z\lesssim 10^{-3.5}Z_{\odot }$. In the hydrogen deuteride cooling
process, the gas temperature would reach that of the cosmic
microwave background (CMB) within a Hubble time.  When the gas
temperature reaches the CMB limit, the Pop III.2 stars with typical
masses $\sim40\ M_{\odot }$ form, which are smaller than the typical
mass of Pop III.1 stars. For simplification, we assume that
the Pop III.2 stars have a lower-mass distribution, but have the
approximate properties of the Pop III.1 stars \citep{ohk09},
including the intrinsic bolometric luminosity, the effective
temperature, the main-sequence lifetime, and the time-averaged
hydrogen photoionization rate. Second, the Pop II stars, as
metal-poor stars with metallicity $Z=1/50\ Z_{\odot }$, form in
clouds that can undergo metal and dust cooling. The clouds are able to
fragment into smaller masses, leading to typical masses less than
those of Pop III stars.

\subsection{Pop III \& Pop II}

We adopt the similar initial mass function (IMF) as same as \citet{coo12} for Pop III.1 and Pop II stars.
The IMF of Pop III.1 stars is \citep{lar98}%
\begin{eqnarray}
f(M_{\ast })\propto M_{\ast }^{-1}\left( 1+\frac{M_{\ast }}{M_{\ast }^{C}}%
\right) ^{-1.35},
\end{eqnarray}%
where $M_{\ast }^{C}=250\ M_{\odot }$, and the mass range is from 5
to 500 $M_{\odot }$.
For Pop II stars, we adopt the IMF given by \citet{sal55}%
\begin{eqnarray}
f(M_{\ast })\propto M_{\ast }^{-2.35},
\end{eqnarray}%
with the mass range from 5 to 150 $M_{\odot }$. As pointed
out by \citet{joh06}, we use an intermediate stellar IMF between Pop
III.1 and Pop II for Pop III.2 stars.
We adopt the IMF given by \citet{lar98} with $M_{\ast }^{C}=150\ M_{\odot }$,
and the mass range is from 5 to 250 $M_{\odot }$. So the mean mass
of Pop III.2 is $\sim40M_{\odot}$, which is consistent with the
typical mass given by \cite{yos07} and \cite{hos11}. The
normalization is given by
\begin{eqnarray}
\int_{M_{\ast ,\min }}^{M_{\ast ,\max }}dM_{\ast }f(M_{\ast })=1.
\end{eqnarray}%
The mean stellar mass of one population is%
\begin{eqnarray}
\overline{M}_{\ast }=\int_{M_{\ast ,\min }}^{M_{\ast ,\max }}dM_{\ast
}M_{\ast }f(M_{\ast }).\label{mmass}
\end{eqnarray}%
Here we use the results from \citet{lej01} and \citet{sch02b} to calculate the
main stellar parameters, such as the intrinsic bolometric luminosity $%
L_{\ast }^{bol}(M_{\ast })$, the effective temperature $T_{\ast
}^{eff}(M_{\ast })$, the main-sequence lifetime $\tau _{\ast }(M_{\ast })$,
and the time-averaged hydrogen photoionization rate $\overline{R}%
_{HI}(M_{\ast })$. At first we define $x=\log _{10}(M_{\ast
}/M_{\odot })$.

For Pop III stars (Pop~III.1 and Pop~III.2), the parameters are
given by
\begin{eqnarray}
\log _{10}(L_{\ast }^{bol}/L_{\odot }) &=&0.4568+3.897x-0.5297x^{2},  \notag
\\
\log _{10}(T_{\ast }^{eff}/\unit{K})
&=&3.639+1.501x-0.5561x^{2}+0.07005x^{3},  \notag \\
\log _{10}(\tau _{\ast }/\unit{yr}) &=&9.785-3.759x+1.413x^{2}-0.186x^{3},
\notag \\
\log _{10}(\overline{R}_{HI}/\unit{s}^{-1}) &=&\left\{
\begin{array}{ll}
39.29+8.55x & (5M_{\odot }\leq M_{\ast }\leq 9M_{\odot }) \\
43.61+4.90x-0.83x^{2} & (9M_{\odot }<M_{\ast }\leq 500M_{\odot }).%
\end{array}%
\right.
\label{popiii}
\end{eqnarray}%
For Pop II stars, the parameters become
\begin{eqnarray}
\log _{10}(L_{\ast }^{bol}/L_{\odot }) &=&0.138+4.28x-0.653x^{2},  \notag \\
\log _{10}(T_{\ast }^{eff}/\unit{K}) &=&3.92+0.704x-0.138x^{2},  \notag \\
\log _{10}(\tau _{\ast }/\unit{yr}) &=&9.59-2.79x+0.63x^{2},  \notag \\
\log _{10}(\overline{R}_{HI}/\unit{s}^{-1})
&=&27.80+30.68x-14.80x^{2}+2.50x^{3}.  \label{popii}
\end{eqnarray}%
This stellar model gives the number of ionizing photons emitted per stellar
baryon
\begin{eqnarray}
\eta _{ion}\approx \left\langle \frac{\overline{R}_{HI}(M_{\ast })\tau
_{eff}(M_{\ast })}{M_{\ast }}\right\rangle \frac{m_{p}}{1-Y},
\end{eqnarray}%
where $m_p$ is the proton mass, $Y=0.25$ is the mass fraction of helium, and $%
\tau _{eff}(M_{\ast })$ is the effective stellar lifetime, given by Eq~(\ref{timeeff}) in Section 4.  The effective stellar lifetime $\tau_{eff}$ may be less
than the real lifetime $\tau _{\ast }$, because some low-mass stars are still not dead.

\subsection{The star formation rate}

Following \citet{gre06}, we consider that the Pop III.1
stars form in the minihalos with the minimum virial temperature $%
T_{vir}\approx 10^{3}\unit{K}$. Pop III.2 stars form in
metal-free halos above $T_{vir}=10^{4}\unit{K}$ via HD cooling.
When star formation begins, a sufficient amount of cold
dense gas accumulating in a dark matter halo is needed. Because of
the collapse of the dark matter halos, the baryonic gas becomes
viral equilibrium with the dynamically-dominant dark matter halos,%
\begin{eqnarray}
\frac{GM_{h}}{R_{vir}}\sim \upsilon _{vir}^{2},
\end{eqnarray}%
where $M_{h}$ is the dark matter halo mass, $R_{vir}$ is the virial radius,
and $\upsilon _{vir}$ is the virial velocity. The virial radius is %
\citep{bar01,bro13}%
\begin{eqnarray}
R_{vir}\approx 0.2\unit{kpc}\left( \frac{M_{h}}{10^{6}M_{\odot }}\right)
^{1/3}\left( \frac{1+z}{10}\right) ^{-1}\left( \frac{\Delta _{c}}{200}%
\right) ^{-1/3},
\end{eqnarray}%
where $\Delta _{c}=\rho _{vir}/\rho _{b}$ is the overdensity after
virialization nearly finishes, $\rho_{vir}$ is the virial density of
dark matter halos, $\rho_b=\Omega_b\rho_{cr}$ is the baryon mass
density of the universe, and $\Delta _{c}\approx 18\pi^{2}$ in the
Einstein--de Sitter model. The gas heats up resulting from the
collapse of the dark matter halo, in which the virial temperature of
the gas corresponds to the virial velocity of dark matter halos. So
$k_{B}T_{vir}\sim \mu m_{p}\upsilon _{vir}^{2}$, which is leading to
\begin{eqnarray}
T_{vir}\approx 10^{3}\unit{K}\,\mu \left( \frac{M_{h}}{10^{6}M_{\odot }}%
\right) ^{2/3}\left( \frac{1+z}{10}\right),
\end{eqnarray}%
where $\mu=1.2,0.6$ is the mean molecular weight for neutral and ionized
primordial gas, respectively. For a dark matter halo with mass of $%
10^{6}M_{\odot }$, the virial temperature in this minihalo is $\sim 10^{3}%
\unit{K}$, which is below the threshold $\sim 10^{4}\unit{K}$ of the cooling
temperature of atomic hydrogen, so that the gas is unable to cool, resulting
in no star formation. The gas would simply persist in hydrostatic
equilibrium. However, the cooling in such a low-temperature primordial gas
could rely on molecular hydrogen (H$_{2}$) instead. The main formation
channel is the sequence: H + e$^{-}$ $\rightarrow $ H$^{-}$ + $\gamma $,
followed by H$^{-}$ + H $\rightarrow $ H$_{2}$ + e$^{-}$ \citep{bro13}. At
last, the evolution of gas inside minihalos, driven by H$_{2}$ cooling,
leads to the formation of Pop III.1 stars with typical masses $\sim
100M_{\odot }$. After the Pop III.1 star formation, the composition of the
primordial gas becomes different, and the cooling process would be more
complicated. The hydrogen deuteride (HD) molecule provides an additional
cooling channel. For such a dark matter halo with mass of $\gtrsim 10^{8}M_{\odot }$%
, the cooling of HD would be efficient with temperature $\sim 10^{4}\unit{K}$%
, leading to the formation of Pop III.2 with the typical masses $\sim 40M_{\odot }$.

In order to obtain the SFRs of the early stars, we consider a
semi-analytic approach by using the collapse fraction function of a dark
matter halo. The famous one is the Press--Schechter formalism providing a
way to calculate the abundance of the mass of dark matter halos %
\citep{pre74}. For a given power spectrum $P\left( k\right) \propto
\left\vert \delta _{k}\right\vert ^{2}$, we adopt the matter power
of \citet{kom09}. The Gaussian variance of the fluctuations on the
mass-scale $M_h$ is%
\begin{eqnarray}
\sigma _{M_{h}}^{2}=\frac{1}{2\pi ^{2}}\int_{0}^{\infty }P\left( k\right)
W\left( k,R_h\right) ,
\end{eqnarray}%
where $M_{h}=\left( 4/3\right) \pi \rho _{vir}R_h^{3}$ and $W\left(
k,R_h\right) $ is a top-hat filter function
\begin{eqnarray}
W=\frac{3}{\left(k R_{h}\right) ^{3}}\left[ \sin \left( k R_{h}\right)
-\left( k R_{h}\right) \cos \left( k R_{h}\right) \right] .
\end{eqnarray}
The comoving number density of dark matter halos per unit mass could be
given by \citet{pre74} formalism
\begin{eqnarray}
\frac{dn_{PS}}{dM_{h}}=\frac{\Omega _{m}\rho _{cr}}{M_h^{2}}\sqrt{\frac{2}{%
\pi }}\frac{\delta _{c}}{\sigma _{M_h}D\left( z\right) }e^{-\delta
_{c}^{2}/2\sigma _{M_h}^{2}D^{2}\left( z\right) }\frac{-d\ln \sigma _{M_{h}}%
}{d\ln M_{h}},  \label{PS}
\end{eqnarray}%
where the critical density of the universe is $\rho
_{cr}=1.8785\times 10^{-29}h^{2}\unit{g}\,\unit{cm}^{-3}$, and
$\delta _{c}=\delta \rho /\rho =1.686$ is the critical overdensity
for a spherical perturbation. For the $\Lambda $CDM model, the
growth factor is given by $D\left( z\right) =g\left( z\right)
/\left( 1+z\right) g\left( 0\right) $, where
\begin{eqnarray}
g(z)=\frac{(5/2)\Omega _{m}(z)}{\Omega _{m}^{4/7}-\Omega _{\Lambda
}+(1+\Omega _{m}(z)/2)(1+\Omega _{\Lambda }(z)/70)},
\end{eqnarray}%
with
\begin{eqnarray}
\Omega _{m}(z)=\frac{\Omega _{m}(1+z)^{3}}{\Omega _{m}(1+z)^{3}+\Omega
_{\Lambda }},
\end{eqnarray}%
\begin{eqnarray}
\Omega _{\Lambda }(z)=\frac{\Omega _{\Lambda }}{\Omega _{m}(1+z)^{3}+\Omega
_{\Lambda }}.
\end{eqnarray}
A fraction of mass in the universe collapsing into halos with the mass more
massive than $M_{h,min}$, referred to as the collapse fraction, is written
as
\begin{eqnarray}
F_{col}=\frac{1}{\rho _{m}}%
\int_{M_{h,min}}^{M_{h,max}}M_{h}n_{PS}(M_{h},z)dM_{h},
\end{eqnarray}%
where $\rho _{m}=\Omega_m\rho_{cr}$ is the matter density of the
universe, and $n_{PS}\left( M_h,z\right)$ is obtained by the PS
formalism. $M_{h,min}$ and $M_{h,max}$ are determined by the virial
temperature. Thus the preliminary SFR is given by \citep{gre06}
\begin{eqnarray}
\psi (z)=f_{\ast }\rho _{m}\frac{\Omega _{b}}{\Omega _{m}}\left\vert \frac{%
dF_{col}}{dz}\right\vert \left\vert \frac{dz}{dt}\right\vert ,
\end{eqnarray}%
where $%
f_{\ast }$ is the star formation efficiency.

The transition from Pop III to Pop II stars is dependent on the
evolution of metallicity. Here, we consider two channel of metal
pollution: supernova-driven protogalactic outflows
\citep{ferr00b,fur05} and the ``genetic channel''
\citep{sch06b,tre09}. At first, we consider that metals are enriched
via supernova-driven protogalactic outflows. For a protogalaxy
within a dark matter halo, metals are blow out by supperbubbles
(SBs) resulting from supernova explosion. According to
\citet{ferr00a}, the condition of blowout is that
the blowout velocity $v_{b} $ should be larger than the escape velocity $%
v_{e}$, which gives the fraction of the mechanical energy of superbubble
that can blowout. The efficiency of the metal escape produced by
SNe is close to unity when blowout does take place \citep{macl99}. If a primordial galaxy has an
exponentially stratified density distribution $\rho \propto \exp \left( {-z/H%
}\right) $, then the shock wave from SN explosion initially is decelerated by
ISM and subsequently accelerated due to blowout in a lower density environment
and continued SB luminosity. The velocity of shock wave has a minimum at $%
z=3H$, which is defined as the blowout velocity. It is given by \citep{ferr00a}
\begin{eqnarray}
v_{b}\approx 2.7\left( \frac{L}{\mu m_{p}nH^{2}}\right) ^{1/3},
\end{eqnarray}%
where $L$ is the mechanical luminosity of SBs, $n$ is the number density of
a uniform ambient medium of protogalaxies, and $H$ is taken to be a free
parameter. For ISM parameters $n=0.5\unit{%
cm}^{3},\,\mu =1.25$, blowout occurs if $v_{b}>v_{e}$. Thus, the
critical mechanical luminosity should be
\begin{eqnarray}
L_{c}=0.05\mu m_{p}nH^{2}\left( \frac{GM_{h}}{R_{h}}\right) ^{3/2}.
\end{eqnarray}%
The total mechanical luminosity $L_t$ must be larger than $L_c$ when
blowout occurs. Here we assume the escape velocity is equal to the
circular velocity of the halo $v_{e}\approx v_{c}=\left(
GM_{h}/R_{h}\right) ^{1/2}$. The total mechanical luminosity of SB
in a galaxy could given by
\citep{ferr98}
\begin{eqnarray}
L_{t}=\varepsilon _{SN}\nu f_{\ast }f_{b}\frac{\Omega _{b}}{\Omega _{m}}%
\frac{M_{h}}{t_{ff}\left( z\right) },  \label{ml}
\end{eqnarray}%
where $\varepsilon _{SN}=10^{51}\unit{erg}$ is the characteristic energy of
a SN, the cooling fraction of baryons $f_{b}\simeq 1$ in the halo $10^{4.3}%
\unit{K}<T_{vir}<10^{5.7}\unit{K}$ \citep{mad01}, and
$f_{\ast }$ is the star formation efficiency. $t_{ff}\left(
z\right) =\left( 4\pi G\rho _{m}\left( z\right) \right) ^{-1/2}$ is
the free fall time of dark matter halo. A
typical mechanical luminosity is $L_{t}\approx 2.3\times 10^{38}\unit{erg}\,%
\unit{s}^{-1}$ for $M_{h}=10^{8}M_{\odot }$ and $z=10$. The mass in stars
per SN event $\nu ^{-1}$, which is given by%
\begin{eqnarray}
\nu ^{-1}=\frac{\int_{M_{\ast ,min}}^{M_{\ast
,max}}M_{\ast }f\left( M_{\ast }\right) dM_{\ast }
}{\int_{8M_{\odot
}}^{M_{\ast ,max}}f\left( M_{\ast }\right) dM_{\ast }}.
\end{eqnarray}%
Here we assume the lowest mass of SN progenitors is $8M_{\odot }$. Eq (\ref%
{ml}) gives the total mechanical luminosity. However, only a
fraction of the mechanical luminosity could blowout from galaxies, because the SNe
might be occur in different regions in one galaxy, and some
superbubbles with low mechanical luminosity could not blowout in certain regions. Thus we calculate the
efficiency of blowout following \citet{ferr00b}. Pop III.1
stars are formed in minihalos, and recent simulation
\citep{bro09} implies that a minihalo only contains one Pop III.1
star. Thus the star formation may be confined to a small region,
leading to SN explosions in the formation of a single superbubble.
In this case, the single superbubble is likely to blowout $L_t>L_c$
and the fraction of the mechanical energy that could blowout is
$\eta_B\simeq1$. For larger galaxies where Pop
III.2 and Pop II stars formed \citep{bro13}, the SNe would occur in
different OB associations, which are more widely distributed within
galaxies. The luminosity function of OB association is approximated
by \citep{oey97,por10}
\begin{eqnarray}
\frac{dN_{OB}}{dL_{OB}}\propto\frac{dN_{OB}}{dN}=KN^{-2},\,(1\leqslant N\leqslant N_{max}),
\end{eqnarray}%
where $N$ is the number of SN in a cluster. The probability to have $N$ SNe in one OB association is $p\approx
N^{-2}$, and the average probability is $\overline{p}\sim 1/N_{max}$. The
average number of SN per OB association $\overline{N_{SN}}\approx
N_{tot}/N_{max}$, The total number of supernovae could be given by $%
N_{tot}\simeq f_{\ast }f_{b}(\Omega _{b}/\Omega
_{m})M_{h}/t_{ff}\left( z\right)$. The total number of OB
associations is $K\sim N_{max}$. The
mechanical luminosity of an OB association is $L_{OB}=N\varepsilon _{SN}/t_{OB}$, where $t_{OB}=40~\unit{Myr}$ is the time at which the lowest mass
SN progenitor expires $\sim 8M_{\odot }$. The total mechanical
luminosity could be given by
\begin{eqnarray}
L_{t}(z)=\int_{1}^{N_{max}}L_{OB}dN_{OB}=N_{max}\frac{\varepsilon _{SN}}{t_{OB}}\ln
N_{max}\simeq N_{tot}\frac{\varepsilon _{SN}}{t_{OB}}.\label{seclum}
\end{eqnarray}%
The first total mechanical luminosity given by Eq(\ref{ml})
is calculated via the star formation rate of a proto-galaxy, which
is related to the mass of the dark matter halo. The second total
mechanical luminosity given by Eq(\ref{seclum}) is to obtain the
efficiency of blowout, which is related to the number of SNe in a
cluster. Physically, the latter mechanical luminosity must be equal
to the former. This relation between the maximum number of SN and
the total number of SN is approximately
\begin{eqnarray}
N_{tot}=N_{max}\ln N_{max}.\label{nn}
\end{eqnarray}%
The above equation describes the dependence of the total number of SN $N_{tot}$ as a function of the maximum number of SN in a cluster, and we
assume that the maximum possible number of SN in a cluster is $%
N_{max}\lesssim 500$, which is consistent with the Monte Carlo
simulation of \citet{ferr00b}. We find that Eq(\ref{nn}) could be
applied to the case of Pop III.1 stars in minihalos with
$M_h\sim10^6M_{\odot}$.   Similarly, the effective mechanical
luminosity leading to blowout is
\begin{eqnarray}
L_{b}(z,>L_{c})=N_{max}\frac{\varepsilon _{SN}}{t_{OB}}\ln \frac{N_{max}}{N_{c}},
\end{eqnarray}
where $N_{c}=L_{c}t_{OB}/\varepsilon _{SN}$ is the number of SN in a cluster
with mechanical luminosity. Thus the fraction of the mechanical energy that
could blowout is defined as \citep{ferr00b}
\begin{eqnarray}
\eta _{B}\equiv\frac{L_{b}}{L_{t}}=\frac{\ln (N_{max}/N_{c})}{\ln N_{max}}.
\end{eqnarray}

After the shock wave of SBs propagates to IGM, the metal bubble
would be enriched to a larger zone due to the lower density of IGM.
The outflow will be confined by the pressure of IGM, which
determines the radius
of the metal bubble. The shell growth stalls when $P_{i}=P_{0}$, where $%
P_{0} $ is the IGM pressure in the surroundings of the galaxy as $%
P_{0}=n(z)k_{B}T$, where $n(z) $ is the average baryon number density of the
cosmology, and the gas of IGM is heated to $T\simeq 2\times 10^{4}\unit{K}$
by photoionization heating. During the ionized zone overlapping, the
SN-driven bubbles would propagate in the photoionization gas. Here we assume
that the outflow propagates in ionized zone.

We consider the standard evolution for an adiabatic, pressure-driven
superbubble. The growth of the SN shell radius is \citep{wea77}
\begin{eqnarray}
R=(\frac{125}{254})^{1/5}(\frac{Lt^{3}}{\mu m_{p}n_{IGM}})^{1/5},
\end{eqnarray}%
where $n_{IGM}\approx n\left( z\right) $ is the number density of the
ambient medium in IGM. The interior pressure is
\begin{eqnarray}
P_{i}=\frac{7}{(3850\pi )^{2/5}}L^{2/5}(\mu m_{p}n_{IGM})^{3/5}t^{-4/5}.
\end{eqnarray}%
When the shell growth stalls $P_{i}=P_{0}$, we have
\begin{eqnarray}
R_{s}=\frac{5\cdot 7^{1/4}}{\left( 550\pi \right) ^{1/2}}L_{e}^{1/2}(\mu
m_{p}n_{IGM})^{1/4}P_{0}^{-3/4},
\end{eqnarray}%
where $L_{e}=\eta _{B}L_{t}$ is the effective mechanical luminosity,
which is the fraction available for blowout into IGM. The stall age
is
\begin{eqnarray}
t_{s}=\frac{7^{3/4}}{\left( 550\pi \right) ^{1/2}}(\mu
m_{p}n_{IGM})^{3/4}P_{0}^{-5/4}L_e^{1/2}.
\end{eqnarray}%
The typical parameters are $R_{s}\approx 33\,\unit{kpc}$ and $t_{s}\approx 1.5\,%
\unit{Gyr}$ for $M_{h}=10^{8}M_{\odot }$ and $z=10$. If $z\gg 1\ (\Omega
_{M}\gg \Omega _{\Lambda })$, the stall redshift is given by \citep{voi96}
\begin{eqnarray}
z_{s}\approx \left( \sqrt{1+z}-\frac{H_{0}t_{st}}{2}\right) ^{2}-1,
\end{eqnarray}%
where $t_{st}=t_{ISM}+t_{s}$. $t_{ISM\text{ }}$is the time during which a SN shell
spreads through a galaxy, which is much less than the stall age in the IGM.
The comoving metal-enrich radius is $R_{e}\approx (1+z_{s})R_{s}$. If $%
t_{s}\lesssim 1/H_{0},$ the fraction of space with metals is approximately
given by
\begin{eqnarray}
Q_{e}^\prime(z)\approx \int_{M_{h,min}}^{\infty }dM_{h} \frac{4\pi }{3}R_{e}^{3}%
\left( M_{h},z_{s}\right) \frac{dn_{PS}}{dM_{h}}(M_{h},z).\
\end{eqnarray}
This equation could be well approximate if the metal-enriched bubble
did not overlap and the expanding time is much less than the Hubble
time $H^{-1}(z)$ . If the protogalaxies are randomly distributed,
then the filling factor would be $p_{e}^\prime(z)=1-\exp
[-Q_{e}^\prime(z)]$. In fact, due to clustering, some new halos form
in the metal-enriched regions. The fraction of space with metals
should be corrected by an excess probability that two galaxies are
located near each other, viz, the galaxy two-point correlation
function $\xi_{gg}(R_e)=b_0b_m\xi_{hh}(R_e)$ \citep{fur05}, where
$\xi_{hh}$ is the correlation function of dark matter halo (e.g.
\citet{gre06}), $b_0\simeq b(M_{h,min})$ is the bias of the newly
formed galaxies and $b_m$ is the bias of the metal enriched regions,
which is given by
\begin{eqnarray}
b_m=\frac{\int dM_{h} (4\pi/3)\rho R_{e}^{3}b(m)
(dn_{PS}/dM_{h})}{\int dM_{h} M_h
(dn_{PS}/dM_{h})}.
\end{eqnarray}
Thus, the probability that a new halo lies within metal-enriched
region is then approximately given by
\begin{eqnarray}
Q_e=Q_e^\prime[1+\xi_{gg}(R_e)]
\end{eqnarray}
Finally, the filling factor would be $p_{e}(z)=1-\exp [-Q_{e}(z)]$
after assuming the wind hosts are randomly.

On the other hand, the metal enrichment of galaxies can proceed via
``genetic channel'' \citep{sch06b}, that is, metals are enriched via
the merger of the lower mass progenitors, rather than through
outflows from neighbours. Here, we calculate the probability
$f_{old}(z)$ that a new collapsing halos accreting onto some old
halos via Extended Press-Schechter model (the Appendix of
\citet{fur05}), which is
\begin{eqnarray}
f_{old}(z)=\frac{\int_{2M_{h,min}}^\infty dM_h M_h(dn_{PS}/dM_h)F(<M_{h,min},z_h|M_h,z)}{\int_{M_{h,min}}^\infty dM_h M_h(dn_{PS}/dM_h)F(<M_{h,min},z_h|M_h,z)},
\end{eqnarray}
where the factor 2 is attributed to the assumption that a parent
halo with a mass $M_h<2M_{h,min}$ would be included in the new halo
component, and the fraction of the accreted mass in a halo with a
mass $M_h$ at redshift $z$ is
\begin{eqnarray}
F(<M_{h,min},z_h|M_h,z)=\mathrm{erf}\left(\frac{\delta_c(z_h)-\delta_c(z)}{\sqrt{2(\sigma_{M_{h,min}}^2-\sigma_{M_h}^2)}}\right),
\end{eqnarray}
where $z_h$ corresponds to some earlier time when a parent halo with
a mass ($<M_{h,min}$) formed, which is fixed by the dynamic time
within a galaxy \citep{fur05}. After a halo merging a mass above
$M_{h,min}$, it would most likely form Pop II stars. Thus, the
probability $p_{gc}(z)$ that a dark matter halo did not inherit any
metals from its progenitor is $p_{gc}(z)\simeq1-f_{old}(z)$.

\begin{figure}
\centering\includegraphics[angle=0,scale=.30]{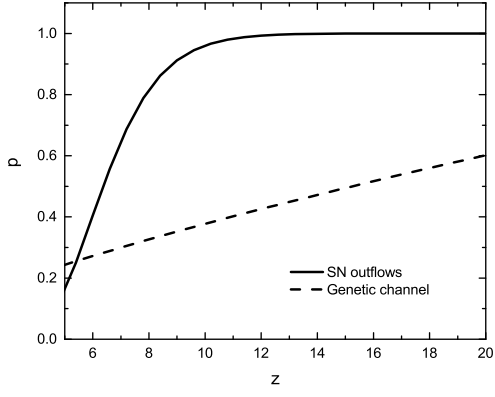}
\caption{The evolution of the non-metal fraction.  The solid curve corresponds to the evolution of the pristine fraction $p_{pris}(z)=1-p_e(z)$, and the dashed curve corresponds to the evolution of $p_{gc}(z)$ that a dark matter halo did not inherit any metals from its progenitor.}
\label{fig1}
\end{figure}

In Figure \ref{fig1}, the solid curve corresponds to the evolution
of the pristine fraction $p_{pris}(z)=1-p_e(z)$, and the dashed
curve corresponds to the evolution of $p_{gc}(z)$. We find that the
metals of IGM begin to be significantly enriched via
supernova-driven protogalactic outflows at $z\sim 10$. However, due
to $p_{gc}(z)\sim 0.2-0.6$ during a large range redshift, the metal
enrichment of galaxies is still dominated by ``genetic channel'' at
$z\gtrsim6$.

Since the Pop III stars generate many ionizing photons,
the ionized bubbles would be photoheated to $\sim 10^{4}\unit{K}$,
which prevent the ionized gas to collapse fresh stars. Thus, the formation of
Pop III.1 stars is suppressed by a factor being equal to the volume filling fraction of ionized regions $%
Q_{ion}(z)$, which is discussed in the next section. The SFR of Pop
III.1 stars is given by
\begin{eqnarray}
\psi _{III.1}(z)&=&f_{\ast ,III.1}\rho _{m}\frac{\Omega _{b}}{\Omega _{m}}%
(1-p_{e}(z))p_{gc}(z)\nonumber\\
&\times&(1-Q_{ion}(z))\left\vert \frac{dF_{coll}}{dz}\right\vert
_{T_{vir}=10^{3}\unit{K}}^{T_{vir}=10^{4}\unit{K}}\left\vert \frac{dz}{dt}%
\right\vert. \label{sfr31}
\end{eqnarray}%
As pointed out by \citet{joh06}, the free electrons can boost the production of $\mathrm{H_{2}}$%
, leading to a lower temperature where the $\mathrm{HD}$ can be
cooled. These stars, originating from metal-free gas and cooling via
$\mathrm{HD}$ channel, would be less massive than Pop III.1 stars.
The formation of Pop III.2 stars requires an increased abundance of
free electrons. Here we consider two main pathways toward Pop III.2
\citep{bro13}%
. The first one results from the photo-ionization of ambient neutral
hydrogen of Pop III.1 stars, and the non-equilibrium recombination
leads to a boosted abundance of $\mathrm{H_{2}}$ and $\mathrm{HD}$
after the Pop III stars have died \citep{yos07}. In order to prevent
pre-enrichment of the gas, these Pop III stars ($260<m<500M_{\odot
}$) have to directly collapse to black holes, the number fraction of
which is $f_{n}=\int_{260M_{\odot}}^{500M_{\odot}}dM_{\ast
}f(M_{\ast })\simeq6\times 10^{-2}$. The second one arises from the
collision ionization in shocks that originate from the collapse of
metal free gas of more massive dark matter halos with
$T_{vir}=10^{4}\unit{K}$ \citep{gre06}. Thus, the SFR of Pop III.2
is
\begin{eqnarray}
\psi _{III.2}(z)&=&\rho _{m}\frac{\Omega _{b}}{\Omega _{m}}(1-p_{e}(z))p_{gc}(z)
(f_{\ast ,III.1}f_{n}Q_{ion}(z)\nonumber\\
&\times &\left\vert \frac{dF_{coll}}{dz}\right\vert
_{T_{vir}=10^{3}\unit{K}}^{T_{vir}=10^{4}\unit{K}}
+f_{\ast ,III.2}\left\vert
\frac{dF_{coll}}{dz}\right\vert _{T_{vir}=10^{4}\unit{K}})\left\vert \frac{dz%
}{dt}\right\vert.\label{sfr32}
\end{eqnarray}%
\newline
In fact, the first term in the above equation is small due to the
number of the stars that directly collapse to black hole is very
few. For Pop II stars, we assume that they are formed from the
metal-enriched dark matter halos with $T_{vir}=10^{4}\unit{K}$. The
SFR of Pop II is
\begin{eqnarray}
\psi _{II}(z)=f_{\ast ,II}\rho _{m}\frac{\Omega _{b}}{\Omega _{m}}%
[1-(1-p_{e}(z))p_{gc}(z)]\left\vert \frac{dF_{coll}}{dz}\right\vert _{T_{vir}=10^{4}\unit{K}%
}\left\vert \frac{dz}{dt}\right\vert.\label{sfr2}
\end{eqnarray}
In Eq(\ref{sfr31}) and Eq(\ref{sfr32}), the SFRs depend on the hydrogen reionization fraction $Q_{ion}$ that is as a
function of redshift $z$, which is discussed in the next section.

Due to inhomogeneous metal pollution, some regions where the metal
enrichment is overlapped have larger metallicity. The upper limit of
metal-enriched IGM metallicity in these rich regions could be
estimated by
\begin{eqnarray}
Z_{rich}\left( z\right) =\frac{y_{II}}{p_e(z)\rho _{b}}%
\int_{z}^{\infty }\eta_B(M_{h,min,z^\prime})\zeta_e(z^\prime)\psi _{II}\left( z^\prime\right) \left\vert \frac{dt}{dz^{\prime }}%
\right\vert dz^{\prime },\label{meta1}
\end{eqnarray}%
where the metal yields are $y_{II}=0.005$ for Pop II, which is
consistent with the proposed values in \citet{gre06}, $\rho_b$ is
the mean baryon mass density of the universe,
$\zeta_e(z)\psi_{II}(z)$ is the SFR of Pop II stars which are
polluted via supernova-driven protogalactic outflows, and
$\zeta_e(z)=p_e(z)/[1-(1-p_e(z)p_{gc}(z))]$. The metals in the
overlapped metal-enriched region are mainly from the Pop II stars
which are polluted via SNe, thus we ignore the first contribution of
metal enrichment via the SNe of Pop III in this overlap region. On
the other hand, we calculate the upper limit of the metallicity of
the galaxies whose all progenitors were polluted by ``genetic
channel'', which is
\begin{eqnarray}
Z_{gc}\left( z\right) =\frac{y_{II}}{\rho _{ISM}}%
\int_{z}^{\infty }\kappa_v^3(1-\eta_B(M_{h,min},z^\prime))\zeta_{gc}(z^\prime)\psi _{II}\left( z^\prime\right) \left\vert \frac{dt}{dz^{\prime }}%
\right\vert dz^{\prime },\label{meta2}
\end{eqnarray}
where $\zeta_{gc}(z)\psi_{II}(z)$ is the SFR of Pop II stars which
are polluted via ``genetic channel'', and
$\zeta_{gc}(z)=[1-p_{gc}(z)]/[1-(1-p_e(z)p_{gc}(z))]$. We assume
that the mean baryon mass density of ISM is
$\rho_{ISM}\simeq10^{-24}\unit{cm}^{-3}$, and the ratio of the mean
distance between two galaxies to the scale of a galaxy is
$\kappa_v\simeq 100$.

\section{THE REIONIZATION}

The volume filling fraction of hydrogen ionized regions $Q^\prime_{ion}$ is \citep{mad98}
\begin{eqnarray}
\frac{dQ^\prime_{ion}}{dt}=\frac{f_{esc}}{n_{H,0}}\sum_i\psi_iq_i-\frac{Q^\prime_{ion}}{\overline{t}_{rec}},  \label{qion}
\end{eqnarray}%
where $i$ represents Pop II, Pop III.2 and Pop III.1, $f_{esc}$ is the escape
fraction of ionizing photons. In fact, due to clustering, the probability that a fresh halo lies within an hydrogen ionized (shown in Eq(\ref{sfr31}) and Eq(\ref{sfr32})) is
\begin{eqnarray}
Q_{ion}=Q^\prime_{ion}[1+\xi_{hh}(z)]
\end{eqnarray}
$\overline{t}_{rec}$ is the volume
averaged recombination time,
which is given by%
\begin{eqnarray}
\overline{t}_{rec}=[C_{HII}(z)\alpha
_{B}^{rec}n_{H,0}(1+z)^{3}(1+Y/4X)]^{-1},
\end{eqnarray}%
where$%
C_{HII}\left( z\right) \equiv \left\langle n_{HII}^{2}\right\rangle
/\left\langle n_{HII}\right\rangle ^{2}$ is the clumping factor of ionized
hydrogen. We use a simple analytic fit of the form $C_{HII}\left( z\right)
=1+9\left[ \left( 1+z\right) /7\right] ^{-2}$ for $z>6$, and $C_{HII}\left(
z\right) =10$ for $z\leq 6$ \citep{gre06}. We assume that the mass fractions of hydrogen and
helium are $X=0.75$ and $Y=0.25$, respectively. Thus the mean hydrogen
number density at $z=0$ is given by
\begin{eqnarray}
n_{H,0}=\frac{X\Omega _{b}\rho _{cr}}{m_{p}}\approx 1.956\times 10^{-7}\unit{%
cm}^{-3}.
\end{eqnarray}%
The number of hydrogen-ionizing photons per stellar mass during
stellar lifetime is
\begin{eqnarray}
q_{i}=\left\langle \frac{\overline{R}_{HI}^{i}\tau _{eff}^{i}}{%
M_{\ast }^{i}}\right\rangle,
\end{eqnarray}%
where $i$ presents Pop II, Pop III.2 and Pop III.1 stars, and $\tau _{eff}(M_{\ast })$ is the effective stellar lifetime, which
may be less than $\tau _{\ast }$, given by Eq. (\ref{timeeff}). The optical
depth for Thomson scattering determined by the ionization history is%
\begin{eqnarray}
\tau =c\sigma _{T}\int_{0}^{\infty }dz^{\prime }Q_{ion}\left( z^{\prime
}\right) n_{H,0}(1+z^{\prime })^{3}(1+Y/4X)\frac{dt}{dz^{\prime }}.
\end{eqnarray}

\begin{figure}
\centering\includegraphics[angle=0,scale=.30]{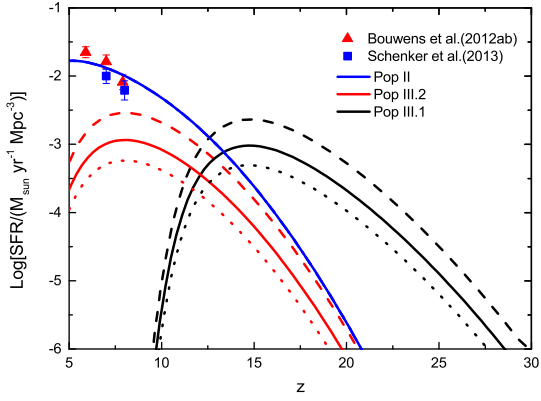} \caption{The three epochs of the cosmic star
formation history: the blue line denotes Pop II stars, the black
line denotes Pop III.1 stars, and the red line denotes Pop III.2 stars.
Case A ($f_{\ast,III}=0.01,f_{esc}=0.21$), Case B ($f_{\ast,III}=0.004,f_{esc}=0.45$) and Case C ($f_{\ast,III}=0.002,f_{esc}=0.75$) are denoted by dashed, solid
and dotted lines, respectively. The observed data is given by \citet{bou12a,bou12b} and \citet{sch13}} \label{fig2}
\end{figure}

In order to constrain our star formation model with fewer
parameters, we adopt a fixed optical depth of Thomson scattering
$\tau=0.08$, which is satisfied with the observation of WMAP
nine-year data with $\tau=0.089\pm0.014$ \citep{hin13} and the
observation of Planck results with $\tau=0.066\pm0.016$
\citep{ade15}. The star formation efficiency of Pop II stars is
assumed to be $f_{\ast,II}=0.01$, which is constrained by the
observation of the SFR at $z\sim 5-10$ \citep{bou12a,bou12b,sch13}.
We take the same value of the star formation efficiency
$f_{\ast,III}$ for both Pop III.1 and Pop III.2 and the same value
of the escape fraction of ionizing photons for all populations
$f_{esc}$. We assume that stars form from redshift $z_{in}=30$.
Figure \ref{fig2} shows the three epochs of the cosmic star
formation history: the blue line denotes Pop II stars, the black
line denotes Pop III.1 stars, and the red line denotes Pop III.2
stars. Because the negative feedback from a star forming in a dark
matter halo prevents the formation of other stars in the same halos,
leading to Pop III stars have a lower star formation efficiency than
Pop II stars, and we set $f_{\ast,III}\leqslant f_{\ast,II}$. For
the same optical depth $\tau=0.08$, we consider three cases: Case A
($f_{\ast,III}=0.01,f_{esc}=0.21$), Case B
($f_{\ast,III}=0.004,f_{esc}=0.45$) and Case C
($f_{\ast,III}=0.002,f_{esc}=0.75$), which are denoted by dashed,
solid and dotted lines, respectively. For the same $\tau$ value, a
larger value of $f_{esc}$ will result in a lower SFR and a lower
value of $f_{esc}$ in a larger SFR. We also find that the SFR of Pop
II raising earlier due to the metal pollution via ``genetic
channel''.

\begin{figure}
\centering\includegraphics[angle=0,scale=.30]{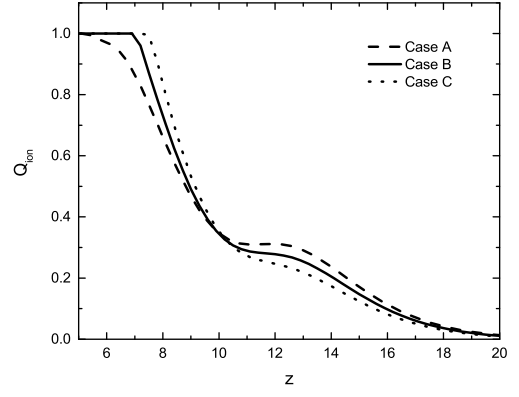} \caption{The
hydrogen reionization fraction $Q_{ion}$ as a function of redshift
$z$. Case A ($f_{\ast,III}=0.01,f_{esc}=0.21$), Case B ($f_{\ast,III}=0.004,f_{esc}=0.45$) and Case C ($f_{\ast,III}=0.002,f_{esc}=0.75$) are denoted by dashed, solid
and dotted lines, respectively.} \label{fig3}
\end{figure}

Figure \ref{fig3} shows the reionization history for three star
formation models. The end of reionization is $z_{end}=5.1,\ 6.9,\
7.2$, corresponding to Case A, B and C, which are denoted by dashed,
solid and dotted lines, respectively. We find that larger SFR of Pop
III.1 stars causes a larger hydrogen reionization fraction at high
redshift, however, lower $f_{esc}$ leads to a later end of
reionization. Due to larger SFR of Pop III.1 at high redshift and a
huge amount of ionizing photons generated by Pop III.1 stars, the
hydrogen reionization fraction $Q_{ion}$ raises at a higher redshift
$z\sim20$ until the SFR of Pop III.1 becoming low at $z\sim10$. Soon
the Pop III.2 and Pop II stars rapidly form, leading to the hydrogen
reionization fraction $Q_{ion}$ raising rapidly until the universe
completely ionized.

\begin{figure}
\centering\includegraphics[angle=0,scale=.30]{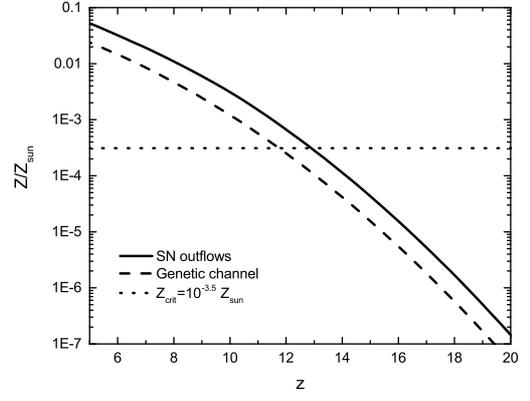}
\caption{The evolution of the upper limit of the metallicity. The soild line denotes the upper limit of the metallicity  in the metal-enriched region of IGM. The dashed line denotes the upper limit of the metallicity in the galaxies whose all progenitors were polluted via ``genetic channel''. The dotted line denotes the critical metallicity $Z_{crit}=10^{-3.5}Z_\odot$.} \label{fig4}
\end{figure}

According to Eq (\ref{meta1}) and Eq (\ref{meta2}), we obtain the
evolution of the upper limit of the metallicities in the
metal-enriched region of IGM and in the galaxies whose all
progenitors were polluted via ``genetic channel'', as shown in
Figure \ref{fig4}. The upper limit is attributed to the
inhomogeneous metal pollution. We predict that: (i) for the
metal-enriched region of IGM, the upper limit of the metallicity
reaches $Z_{crit}=10^{-3.5}Z_{\odot }$ \citep{bro01,sch02a}$\ $at $%
z\sim13$; (ii) for the galaxies polluted via ``genetic channel'', it
reaches the critical value at $z\sim11$. Pop II stars with low mass
formed for $Z\gtrsim Z_{crit}$, which is in good agreement with the
star formation model in Figure \ref{fig2}. Note that as shown in
Figure \ref{fig1}, the metal enrichment is dominated by ``genetic
channel'', however, at a certain redshift, the upper limit of the
metallicity of the metal-enriched region is larger than that of the
galaxies polluted via ``genetic channel''. The reason is that the
fraction of space with metals is so small that the metallicity would
be large, if the Pop II stars that were polluted via SN outflows
always form in this metal-enriched region.

\section{THE NEAR-INFRARED BACKGROUND}

The intensity of the NIRB offers a window of probing the era of
reionization. Following \citet{fer06}, we calculate the NIRB from the epoch
of reionization. There are several contributions to the emission of NIRB:
the continuum emission from stars themselves $\overline{L}_{\nu }^{\ast }$,
the series of recombination lines $\overline{L}_{\nu }^{line}$, the
free-free and free-bound continuum emission from the ionized gas or nebula $%
\overline{L}_{\nu }^{cont}$, and the two-photon emission $\overline{L}_{\nu
}^{2\gamma }$. For the first stars, one of the remarkable properties is a
\textquotedblleft bump\textquotedblright\ in the spectrum of NIRB, which is
from the Lyman $\alpha $ emission \citep{fer13b}. The Lyman $\alpha $ bump
would be higher if Pop III stars were more massive and presented at lower
redshifts, and it would evolve with the transition from Pop III stars to Pop
II stars. For the IGM, the hydrogen density is lower than that of
the stellar nebulae. Here we neglect the emission from IGM, because it is
only a small part of NIRB \citep{coo12}. The intensity of the NIRB is given
by
\begin{eqnarray}
I_{\nu }=\frac{c}{4\pi }\int dz\frac{p((1+z)\nu ,z)}{H(z)(1+z)},
\end{eqnarray}%
where the volume emissivity is given by
\begin{eqnarray}
p(\nu ,z)=\frac{\psi (z)}{\overline{M}_{\ast }}\underset{\alpha
}{\sum }\int dM_{\ast }f(M_{\ast })\overline{L}_{\nu }^{\alpha
}(M_{\ast })\tau _{eff}(M_{\ast }).
\end{eqnarray}%
The stellar effective lifetime could be approximately given by%
\begin{eqnarray}
\tau _{eff}(M_{\ast })=\min [\tau _{\ast }(M_{\ast }),T_{s}\left( z\right) ].
\label{timeeff}
\end{eqnarray}%
$%
\overline{L}_{\nu }^{\alpha }(M_{\ast })$ is the time-averaged luminosity in
the frequency interval $d\nu $ for a radiative process $\alpha $ for stellar
or nebular component, which consists of the stellar blackbody emission, as
well as the reprocessed nebular emission, such as two-photon emission,
recombination line, free-free and free-bound continuum emission. $T_{s}$ is
the time from the formation of the first stars of the universe to the age of
the universe at redshift $z$ \citep{fer06}.

Following \citet{fer06}, we calculate four contributions to the emission of
NIRB: (1). The stellar spectrum, which is the Planck function with the Lyman
absorption:%
\begin{eqnarray}
\overline{L}_{\nu }^{\ast }(M_{\ast })=\left\{
\begin{array}{ll}
4\pi R_{\ast }^{2}(M_{\ast })B_{\nu }(T_{\ast }^{eff}(M_{\ast })), & h\nu
<13.6\,\unit{eV}, \\
0, & h\nu \geq 13.6\,\unit{eV},%
\end{array}%
\right.
\end{eqnarray}%
where $B_{\nu }\left( T_{eff}\right) $ is the Planck spectrum. $R_{\ast }$
is the stellar radius which is determined by the intrinsic bolometric
luminosity $L_{\ast }^{bol}\left( M_{\ast }\right) $ and the effective
temperature $T_{\ast }^{eff}\left( M_{\ast }\right) $ are given by Eq(\ref%
{popiii}) and Eq(\ref{popii}), respectively; (2). The luminosity of
two-photon emission, which is given by%
\begin{eqnarray}
\overline{L}_{\nu }^{2\gamma }(M_{\ast })=\frac{2h\nu }{\nu _{Ly\alpha }}%
(1-f_{Ly\alpha })P(\nu /\nu _{Ly\alpha })\overline{R}_{HI}(M_{\ast }),
\end{eqnarray}%
where $\nu _{Ly\alpha }=2465\unit{THz}$, $f_{Ly\alpha }=0.64$, $P\left(
y\right) dy$ is normalized probability of generating one photon via
two-photon decay in the range $dy=d\nu /\nu _{Ly\alpha }$ for $y\equiv \nu
/\nu _{Ly\alpha }<1$ \citep{fer06}; (3). The line luminosity, which is given
by%
\begin{eqnarray}
\overline{L}_{\nu }^{line}(M_{\ast })=f_{Ly\alpha }h\nu _{Ly\alpha }\phi
(\nu -\nu _{Ly\alpha })\overline{R}_{HI}(M_{\ast }),
\end{eqnarray}%
where $\phi (\nu -\nu _{Ly\alpha })$ is the line profile, which is taken to
be a $\delta $-function $\phi (\nu -\nu _{Ly\alpha })=\delta ^{D}(\nu -\nu
_{Ly\alpha })$. Note that the above equation is for the Lyman alpha line only. (4). The free-free and free-bound continuum luminosity, that
is%
\begin{eqnarray}
\overline{L}_{\nu }^{cont}(M_{\ast })&\simeq& 6.8\times 10^{-38}\frac{%
T_{g}^{-1/2}}{\alpha _{B}^{rec}}\overline{R}_{HI}(M_{\ast })\phi
_{2}(T_{g})e^{-h\nu /kT_{g}}\nonumber\\
&\times&\left[ \overline{g}_{ff}+\frac{R_{y}}{kT_{g}}%
\sum_{n=2}^{\infty }\frac{e^{R_{y}/(kT_{g}n^{2})}}{n^{3}}g_{fb}(n)\right] ,
\end{eqnarray}%
where the case B recombination coefficient is $\alpha _{B}^{rec}\simeq
2.17\times 10^{-10}T_{g}^{-0.7395}$, $\overline{g}_{ff}$and $g_{fb}(n)$ are
the Gaunt factors for free-free and free-bound emission respectively,
approximately, $\overline{g}_{ff}\approx 1.1$ and $g_{fb}(n)\approx 1.05$. $%
T_{g}$ is the gas temperature, here we assume that $T_{g}\approx 10^{4}\unit{%
K}$, and the line profile is $\phi _{2}(10^{4}\unit{K})\approx 1.0$. Then $%
\overline{R}_{HI}(M_{\ast })$ is given by Eq(\ref{popiii}) and Eq(\ref{popii}%
).

\begin{figure}
\centering\includegraphics[angle=0,scale=.30]{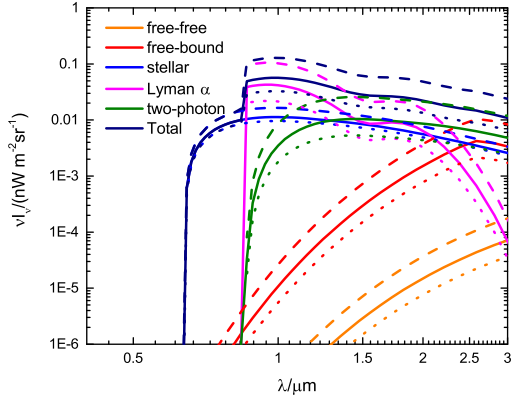} \caption{The spectra of NIRB at redshift $z>6$. Case A ($f_{\ast,III}=0.01,f_{esc}=0.21$), Case B ($f_{\ast,III}=0.004,f_{esc}=0.45$) and Case C ($f_{\ast,III}=0.002,f_{esc}=0.75$) are denoted by dashed, solid and
dotted lines, respectively. Different color lines denote different
contributions to the emission of NIRB.} \label{fig5}
\end{figure}

\begin{figure}
\centering
\includegraphics[angle=0,scale=.30]{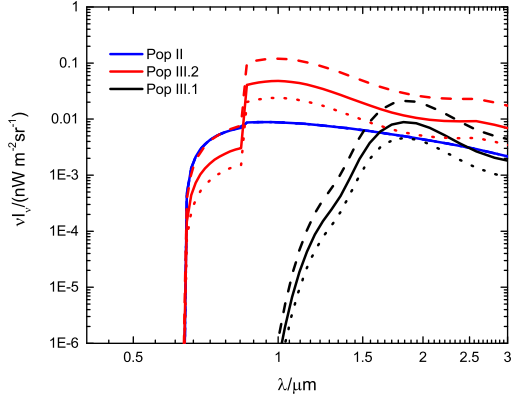}
\caption{The contributions of the different stellar populations to NIRB. The blue, black and red lines denote the NIRB contribution of Pop II, Pop III.1 and Pop III.2, respectively. Case B ($f_{\ast,III}=0.004,f_{esc}=0.45$) and Case C ($f_{\ast,III}=0.002,f_{esc}=0.75$) are denoted by dashed, solid and
dotted lines, respectively.} \label{fig6}
\end{figure}

Figure \ref{fig5} shows the spectra of NIRB at $z>6$. Case A, B and
C are denoted by dashed, solid, dotted lines, respectively.
Different color lines denote different contributions to the emission
of NIRB. The Lyman $\alpha $ and two-photons emission are dominant.
We find that larger value of  $f_{\ast,III}$ leads to larger Lyman
$\alpha$ bump and stronger two-photon emission. The mean intensity
of NIRB contributed by these early
stars during reionization is nearly $\sim 0.03-0.2~\unit{nW}\unit{m}^{-2}\unit{%
sr}^{-1}$ at $z>6$. We also plot the contributions of the different
stellar populations to NIRB, as shown in Figure \ref{fig6}. The
blue, black and red lines denote the NIRB contribution of Pop II,
Pop III.1 and Pop III.2, respectively. We find that the component of
the Pop III.2 dominates the total NIRB spectra and the two Lyman
$\alpha$ bumps ($\sim1215(1+z)\AA$) associated to Pop III.2 and Pop
III.1 are obvious at $z>6$. If we can detect the Lyman $\alpha$
bumps, it would reveal many properties of the early stars, such as
the star formation rates, redshift distribution and the
classification of the Pop III stars.

\section{STOCHASTIC BACKGROUND OF GRAVITATIONAL WAVES}

In this section, we calculate the SBGWs that are generated by these
stars collapse to black holes. The SBGW is mainly dependent on the
SFR and IMF of Pop III and Pop II stars. The flux received
in gravitational waves is
\begin{eqnarray}
F_{\nu }(\nu _{obs})=\int \frac{1}{4\pi d_{L}^{2}}\frac{dE_{GW}}{d\nu }\frac{%
d\nu }{d\nu _{obs}}\psi (z) \frac{f(M_{\ast })}{\overline{M}_{\ast }}dM_{\ast }dV.
\end{eqnarray}%
where $d_{L}$ is the luminosity distance, $dE_{GW}/d\nu $ is the
specific energy of the source, $\psi (z)$ is the star formation
rate, $f(M_{\ast })$ is the IMF of one population, and
$\overline{M}_{\ast }$ is the mean stellar mass given by
Eq(\ref{mmass}). As pointed out by \citet{car80}, the specific
energy flux per frequency is
\begin{eqnarray}
f_{\nu }(\nu _{obs})\equiv \frac{1}{4\pi d_{L}^{2}}\frac{dE_{GW}}{d\nu }\frac{%
d\nu }{d\nu _{obs}}=\frac{\pi c^{3}}{2G}h_{BH}^{2},
\end{eqnarray}%
where $h_{BH}$ is the dimensionless gravitational wave
amplitude. The flux of gravitation waves would be
\begin{eqnarray}
F_{\nu }(\nu _{obs})=\frac{\pi c^{3}}{2G}h_{BG}^{2}\nu _{obs}.
\end{eqnarray}%
The integral dimensionless gravitational wave amplitude
produced from all events that stars collapse to black holes is given
by \citep{per10}
\begin{eqnarray}
h_{BG}^{2}=\frac{1}{\nu _{obs}}\int h_{BH}^{2}\psi (z)\frac{f(M_{\ast })}{\overline{M}_{\ast }}dM_{\ast }dV.
\end{eqnarray}%
The comoving volume can be expressed as
\begin{eqnarray}
dV=4\pi d_{C}^{2}\left( \frac{c}{H_{0}}\right) \left[ \Omega
_{m}\left( 1+z\right) ^{3}+\Omega _{\Lambda }\right] ^{-1/2}dz.
\end{eqnarray}%
For the case that a star collapse to a black hole, we assume that
gravitational waves radiate with an efficiency $\epsilon
_{GW}=\Delta E_{GW}/m_{r}c^{2}$, where $\Delta E_{GW}$ is the total
gravitational waves energy and $m_r$ is the mass of black hole. The
efficiency $\epsilon
_{GW}\lesssim 7\times 10^{-4}$, if the collapse is axisymmetric \citep{sta85}%
. The characteristic amplitude is given by \citep{thor87}%
\begin{eqnarray}
h_{BH}\simeq 7.4\times 10^{-20}\epsilon _{GW}^{1/2}(\frac{m_{r}}{M_{\odot }}%
)(\frac{d_{L}}{1\unit{Mpc}})^{-1},
\end{eqnarray}%
and the observed gravitation waves frequency is
\begin{eqnarray}
\nu _{obs}\approx 1.3\times 10^{4}\unit{Hz}(\frac{M_{\odot }}{m_{r}}%
)(1+z)^{-1}.\label{fregw}
\end{eqnarray}%
We consider that the black holes are formed with $M_{\ast
}>25M_{\odot }$ for Pop II and Pop III. For the stars with
$25<M_{\ast }<140M_{\odot }$, the black holes have the same mass of
the helium core of their progenitors \citep{heg02}
\begin{eqnarray}
m_{r}=m_{He}=\frac{13}{24}(M_{\ast }-20M_{\odot }).
\end{eqnarray}%
In the range $140<M_{\ast }<260M_{\odot }$, the stars are completely
disrupted in PISN  explosions, leading
to no black holes, $m_{r}=0$. For $M_{\ast
}>260M_{\odot }$, these stars would directly collapse to black
holes, thus we neglect stellar mass
loss and the masses of the black holes are equal to those of the their progenitor stars, $%
m_{r}=M_{\ast }$. According to Eq~(\ref{fregw}), the range
of frequency is determined by the mass range of IMF after assuming
that the largest redshift of star formation is $z=30$. For Pop II,
the frequency is $\nu _{obs}>6.4\unit{Hz}$; for Pop III.1, the
background is shifts to lower frequencies due to the direct collapse
for $M_{\ast }>260M_{\odot }$, $\nu _{obs}>0.6~\unit{Hz}$; and for
Pop III.2, $\nu _{obs}>6.4\unit{Hz}$, which has the same range as
Pop II stars, because the stars with $140<M_{\ast }<250M_{\odot }$
were disrupted by PISN.

\begin{figure}
\centering\includegraphics[angle=0,scale=.30]{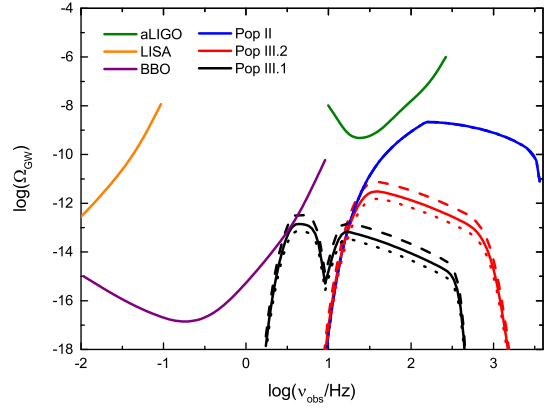} \caption{The
spectrum of the gravitational waves. Pop II, Pop III.2 and Pop III.1
stars are denoted by blue, red, and black lines, respectively. Case B ($f_{\ast,III}=0.004,f_{esc}=0.45$) and Case C ($f_{\ast,III}=0.002,f_{esc}=0.75$) are denoted by dashed, solid and
dotted lines, respectively. The sensitivity
curves of advanced LIGO H1L1, LISA and BBO are denoted by green,
orange and purple, respectively \citep{thr13}, assuming $T=1\,\unit{yr}$ of observation.} \label{fig7}
\end{figure}

The gravitational energy density parameter $\Omega _{GW}$
is defined as the closure energy density per logarithmic frequency
span
\citep{per10}
\begin{eqnarray}
\Omega _{GW}\equiv \frac{1}{\rho _{cr}}\frac{d\rho _{GW}}{d\ln \nu _{obs}}=%
\frac{4\pi ^{2}}{3H_{0}^{2}}\nu
_{obs}^{2}h_{BG}^{2}.
\end{eqnarray}%
In Figure \ref{fig7}, Pop II, Pop III.2 and Pop III.1 stars are
denoted by blue, red, black lines, respectively. The sensitivity
curves of advanced LIGO H1L1, LISA and BBO are denoted by green,
orange and purple, respectively \citep{thr13}, assuming
$T=1\,\unit{yr}$ of the observation. Case A, B and C are denoted by
dashed, solid, dotted lines, respectively. The predicted
gravitational wave background
has a peak amplitude of $\Omega _{GW}\simeq 8\times 10^{-9}$ at $\nu =158%
~\unit{Hz}$ for Pop II star remnants. However, the background
generated by Pop III.2 stars is much weaker than Pop II stars, with a peak amplitude of $%
\Omega _{GW}\simeq 1.2\times 10^{-11}$ at $\nu =28\unit{Hz}$. The
background of Pop III.1 shifted to lower frequencies, and the
amplitude of $\Omega _{GW}$ for Pop III.1 stars shows a minimum
value at $\nu _{obs}\simeq 10\unit{Hz}$,
due to the lack of gravitational wave signals from the stars with $%
140<M_{\ast }<260M_{\odot }$. As shown in Figure \ref{fig7}, it is
difficult to observe the SBGWs from the early stars for the
gravitational wave detectors. However, as shown in Case A, we might
have an opportunity to detect the SBGWs signal from Pop III.1 stars
at observed frequency $\nu_{obs}\simeq3.5\,\unit{Hz}$, which are
twice than the sensitivity of BBO detector.

\section{DISCUSSIONS AND CONCLUSIONS}

In this paper, we have constructed the star formation history for
Pop III and Pop II via the collapse function of dark matter halos.
These SFRs, calculated by the collapse fraction function, are
self-consistent with the observations of reionization (the
reionization optical depth measured by WMAP and Planck, reionization
redshift in range from 5 to 10) and NIRB (the total luminosity
$1\sim 10\,\unit{nW}\unit{m}^{-2}\unit{sr}^{-1}$). At first, Pop III
stars form from pristine baryonic gas in dark matter halos. Due to
the metal enrichment raising in the universe, more and more Pop II
stars formed in metal-enriched region. There are two main channels
of the metal pollution: the supernova-driven protogalactic outflows
\citep{ferr00b,fur05} and the``genetic channel''
\citep{sch06b,tre09}. The former is assumed that the new dark matter
halos are formed in the IGM that is enriched with the metals
dispersed by the first SNe and stellar winds, and the latter
suggests that the new galaxies inherit metals from the lower mass
progenitor galaxies. After the metals in the IGM enriching to a
critical threshold, the Pop I/II stars form gradually. Therefore,
the SFRs of these populations would depend on the hydrogen
reionization fraction and pristine fraction, which are shown in
Eq~(\ref{sfr31}), Eq~(\ref{sfr32}) and Eq~(\ref{sfr2}). Our results
show that the metal enrichment is dominated by ``genetic channel''
during a large redshift range, leading to the earlier raising of the
SFR of Pop II stars, which agrees with the result of \citet{tre09}.
However, the upper limit of the metallicity of the metal-enriched
region could be larger than that of the galaxies polluted via
``genetic channel'' due to the inhomogeneous metal pollution of the
supernova-driven protogalactic outflows.

For the case of the supernova-driven protogalactic outflow, we make a more reasonable assumption
that the SN winds would stall if their pressures are equal to the
pressure of IGM, which is different from the previous papers \citep{fur05,gre06} that calculated the process of the propagation and
distribution of metals by assuming that the SN wind propagated for
half of the age of the universe via Sedov solution \citep{fur05}, however,
Sedov solution is not well approximated after the SN winds have been
stalled by IGM.  Our result shows that the IGM is enriched via SN outflows
during $z=4\sim 10$.

In this paper, we study the NIRB, the SFRs of different populations, and
the reionization history simultaneously, which is different from
\citet{fer06} and \citet{coo12}. In previous works, e.g. \citet{coo12} and \citet{fer13b},
the transition of Pop III to Pop II stars is given by $f_p=(1/2)\{1+\mathrm{erf}[(z-z_t)/\sigma_p]\}$,
where $f_p$ is the fraction of Pop III stars, $z_t$ is the
transition redshift and $\sigma_p$ is the length of the transition. The above equation describing the transition of
Pop III to Pop II might be too simple. In fact, the transfer from Pop III to Pop II stars is a very complex process. Here, we consider the some main properties evolution of three stellar populations, such as the metallicity, reionization and star formation rate. Finally, these properties would affect the character of the spectra of NIRB.

As pointed out by \citet{fer13b}, the Ly$\alpha $ bump could reveal
information of the Pop III era. Our results show that the Pop III
would make a main contribution to the NIRB at high redshift and the
spectra of the NIRB might show two bumps due to the different
distributions of Pop III.1 and Pop III.2 stars.
However, the predicted intensity ($\lesssim0.2~\unit{nW}\unit{m}^{-2}\unit{%
sr}^{-1}$) of NIRB from high redshift ($z>6$) stars is much lower than the contribution of the foreground ($\sim10~\unit{nW}\unit{m}^{-2}\unit{%
sr}^{-1}$), leading to measuring the signal from high-redshift stars
is very difficult. The contributions from low-redshift galaxies and
the bright zodiacal foreground should be correctly subtracted. Many
works have attempted to measure the excess of NIRB without the
contributions of the low-redshift galaxies and other foregrounds
\citep{dwe98,gor00,kas00,tot01,wri01,kas02,
mag03,sal03,coo04a,kas04,kas05a,mat05,kas07b,
tho07a,tho07b,coo12b,kas12}. On the other hand, \citet{fer13b}
suggested that we can detect the relative change in the intensity of
NIRB, which results from a Lyman $\alpha$ bump as a function of
wavelength. However, there are still many ways causing such a
change, such as the evolution of the escape fraction or a rapidly
changing star formation rate as a function of redshift.

We also calculate the stochastic background of gravitational waves
from the collapse of the early stars. However, theoretically, there
are several astrophysical sources contributing to the background of
gravitational waves, including transient sources (e.g. compact
binary coalescence, supernovae, gamma-ray bursts etc.), long-lasting
transient sources (e.g. magnetars, long gamma-ray bursts etc.) and
continuous sources (e.g. pulsar). Thus it is necessary to
distinguish between different sources from the spectrum of
gravitational waves. In this paper, we find that it is difficult to
observe the SBGWs signals from the early stars, however, for
$f_{\ast,III}\simeq0.01$ and $f_{esc}\simeq0.21$,  BBO operating in
the range $0.01-10\unit{Hz}$ might detect lower frequency signal
from Pop III.1 stars, which are twice than the sensitivity of BBO.

\section*{Acknowledgements}
We thank an anonymous referee for valuable and detailed suggestions that have allowed us to improve this manuscript significantly.
This work is supported by the National Basic Research Program of
China (973 Program, grant No. 2014CB845800) and the National Natural
Science Foundation of China (grants 11422325, 11373022, and
11033002), the Excellent Youth Foundation of Jiangsu Province
(BK20140016), and the Program for New Century Excellent Talents in
University (grant No. NCET-13-0279).

\clearpage

\clearpage

\end{document}